\documentclass[conference]{IEEEtran}
\usepackage[final]{graphicx}
\usepackage{subfigure}
\usepackage{float}
\usepackage{amsmath}
\usepackage{cases}
\usepackage{color}
\usepackage{colortbl}
\usepackage{authblk}
\usepackage{algorithm}
\usepackage{algpseudocode}
\usepackage{amsmath}
\usepackage{amssymb}
\usepackage{booktabs}
\usepackage{setspace}
\usepackage{array}
\makeatletter
\renewcommand{\maketag@@@}[1]{\hbox{\m@th\normalsize\normalfont#1}}%
\makeatother
\usepackage{stfloats}
\usepackage{cite}
\usepackage{makecell}
\usepackage{multirow}

\def\BibTeX{{\rm B\kern-.05em{\sc i\kern-.025em b}\kern-.08em
    T\kern-.1667em\lower.7ex\hbox{E}\kern-.125emX}}
    
\ifCLASSINFOpdf
\else
\fi

\usepackage{caption}
\usepackage{mathtools}

\begin{document}
\title{Target Localization with Macro and Micro \\ Base Stations Cooperative Sensing}
\author[1]{Haotian Liu}
\author[1,*]{Zhiqing Wei}
\author[2]{Furong Yang}
\author[1]{Huici Wu}
\author[3]{Kaifeng Han}
\author[1]{Zhiyong Feng}
\affil[1]{Beijing University of Posts and Telecommunications, Beijing 100876, China}
\affil[2]{King's College London, London, United Kingdom}
\affil[3]{China Academy of Information and Communications Technology, Beijing 100191, China}
\affil[1]{Email:\{haotian\_liu, weizhiqing, dailywu, fengzy\} @bupt.edu.cn}
\affil[2]{Email: k2313182@kcl.ac.uk}
\affil[3]{Email: hankaifeng@caict.ac.cn}

\maketitle

\begin{abstract} 
Addressing the communication and sensing demands of sixth-generation (6G) mobile communication system, integrated sensing and communication (ISAC) has garnered traction in academia and industry. With the sensing limitation of single base station (BS), multi-BS cooperative sensing is regarded as a promising solution. 
The coexistence and overlapped coverage of macro BS (MBS) and micro BS (MiBS) are common in the development of 6G, making the cooperative sensing between MBS and MiBS feasible. Since MBS and MiBS work in low and high frequency bands, respectively,
the challenges of MBS and MiBS cooperative sensing lie in the fusion method of the sensing information in high and low-frequency bands. To this end, this paper introduces a symbol-level fusion method and a grid-based three-dimensional discrete Fourier transform (3D-GDFT) algorithm to achieve precise localization of multiple targets with limited resources. Simulation results demonstrate that the proposed MBS and MiBS cooperative sensing scheme outperforms traditional single BS (MBS/MiBS) sensing scheme, showcasing superior sensing performance.
\end{abstract}
\begin{IEEEkeywords}
Cooperative passive sensing,
integrated sensing and communication (ISAC),
high and low-frequency bands, 
macro base station (MBS),
micro BS (MiBS),
multiple targets localization,
signal processing.
\end{IEEEkeywords}

\IEEEpeerreviewmaketitle

\section{Introduction}
Integrated sensing and communication (ISAC), as one of the six key technologies of sixth-generation (6G), has received enormous attention from academia and industry~\cite{wei2023multiple}. 
The emerging applications in 6G mobile communication system, such as internet of vehicles (IoV) and smart transportation, require high-accuracy target localization, which can be satisfied by multi-base station (BS) cooperation~\cite{amhaz2023integrated}. 

In future mobile communication networks, the coexistence and overlapped coverage of macro BS (MBS) and micro BS (MiBS) make their cooperation feasible~\cite{niu2024interference}. With the development of ISAC empowering mobile communication system,
MBS and MiBS cooperative sensing is expected to achieve large-coverage and high-accuracy sensing~\cite{wei2024deep}.
The types of MBS and MiBS cooperative sensing consist of active, passive, and active-passive sensing~\cite{wei2024deep}.
Among the three
types of cooperative sensing, MBS and MiBS cooperative passive
sensing holds enormous research potential due to its advantages of low hardware modification cost and large sensing coverage~\cite{xianrong2020research}.

However, there is limited exploration on the fusion sensing method for target localization with MBS and MiBS cooperative passive sensing, which can be categorized into symbol-level fusion and data-level fusion.
In terms of data-level fusion,
Liu \textit{et al.} in~\cite{liu20226g} proposed a data-level fusion sensing method that the MIBS and MBS individually process received
echo sensing signals, transmitting the sensing information to
a fusion center for localization. This approach reduces the
computational burden on the fusion center but suffers from
a low-accuracy performance of localization. 
In terms of symbol-level fusion, Wei \textit{et al.} in~\cite{wei2023symbol} proposed a symbol-level fusion sensing method in multi-BS cooperative active sensing scenario, which involves the fusion of multiple sensing symbols to achieve high-accuracy sensing. However, this method does not take into account angle information. 

To this end, this paper proposes a symbol-level fusion-enabled high-accuracy MBS and MiBS cooperative passive sensing scheme, which consists of a passive sensing data fusion stage and a sensing processing stage.
In the data fusion stage of passive sensing, preprocessing and symbol-level fusion are separately applied to the echo signals in passive sensing  received by MBS and MiBS to fully utilize the echo information. 
In the sensing processing stage, we propose a grid-based three-dimensional discrete Fourier transform (3D-GDFT) method to achieve high-accuracy localization in the scenarios with limited resources. The simulation verifies that the proposed symbol-level fusion enabled MBS and MiBS cooperative passive sensing scheme utilizes the echo sensing information received by MBS and MiBS to achieve high-accuracy multi-target localization.

The remainder of this paper is structured as follows. Section \ref{se2} outlines the system model. In Section \ref{se3}, we present a symbol-level fusion-enabled MBS and MiBS cooperative passive sensing scheme. Section \ref{se4} details the simulation results, and Section \ref{se5} summarizes this paper.

\section{System Model}\label{se2}
\begin{figure}[!htbp]
    \centering
    \includegraphics[width=0.45\textwidth]{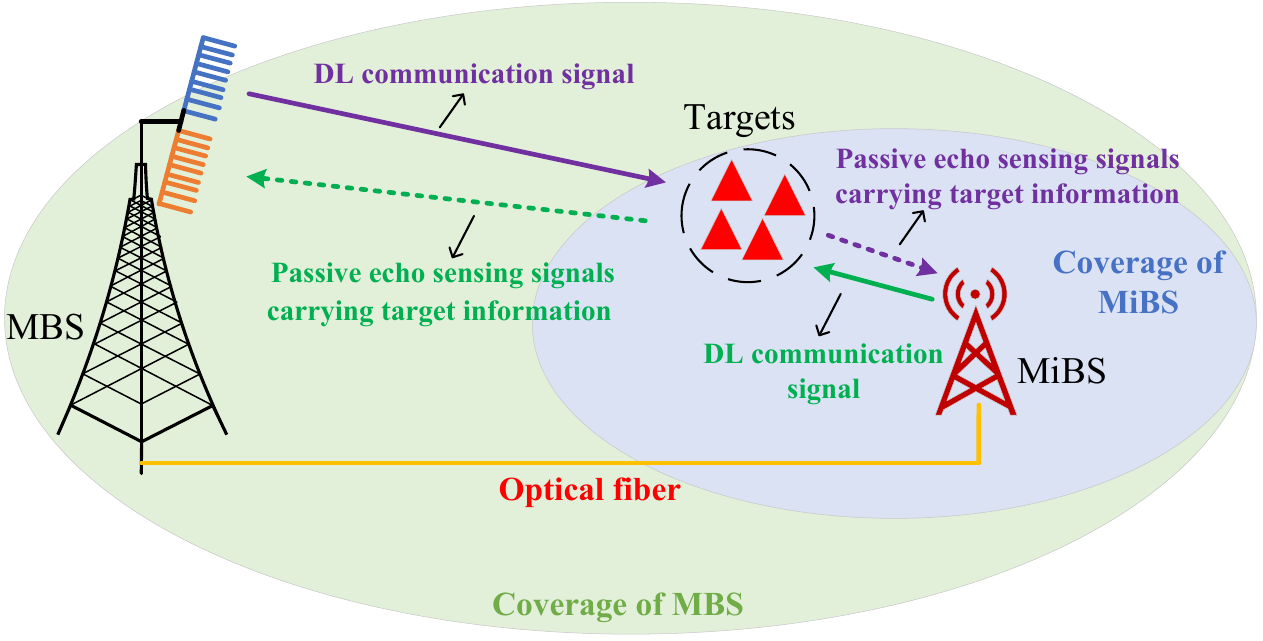}
    \caption{MBS and MiBS cooperative passive sensing}
    \label{fig1}
\end{figure}

We consider MBS and MiBS cooperative passive sensing for target localization in IoV scenario as shown in Fig. \ref{fig1}. 
The ISAC-enabled MBS and MiBS receive and preprocess the echo sensing signals reflected by $L$ targets from the other BS. 
Then, the results are transported to fusion center to estimate the locations of multiple targets.
The MBS and MiBS are implemented with uniform linear arrays (ULAs), which include $N_\text{T}$ transmit antennas and $N_\text{R}$ receive antennas for downlink (DL) and uplink (UL) communication while achieving passive sensing. Meanwhile, MBS and MiBS are connected with optical fiber for sharing information.
\subsection{BS Transmit Signal}
The transmit signals of MBS and MiBS adopt orthogonal frequency division multiplexing (OFDM) signals, which are generally defined as~\cite{liu2023isac}
\begin{equation}\label{eq1}
    s^i(t)=\sum_{m=0}^{M_\text{sym}^i-1}\sum_{n=0}^{N_\text{c}^i-1}\left[
    \begin{array}{l}
\sqrt{P_{m,n,\text{T}}^i}d_{n,m}^i e^{j2\pi(f_\text{c}^i+n\Delta f^i)t} \\ \cdot\text{rect}\left(\frac{t-mT^i}{T^i}\right)
    \end{array}\right],
\end{equation}
where $i$ = 1 or 2 represents the signals from MBS or MiBS; $M_\text{sym}$ and $N_\text{c}$ are the number of OFDM symbols and subcarriers, respectively; 
$P_{m,n,\text{T}}^i$ denotes the transmit power of each resource element (RE); 
$d_{n,m}^i$ is the modulation symbol of the $n$-th subcarrier during the $m$-th OFDM symbol time;
$f_\text{c}^i$ and $\Delta f^i$ stand for the carrier frequency and subcarrier spacing, respectively;
$T^i=\frac{1}{\Delta f^i}+T_\text{cp}^i$ is the total OFDM symbol duration with $T_\text{cp}^i$ is the guard interval~\cite{chen2024downlink}.
$Q=\frac{\Delta f^2}{\Delta f^1}$ denotes the ratio of subcarrier spacing between high and low frequency bands. 

\subsection{BS Received Echo Sensing Signal}
According to Fig.~\ref{fig1}, the received echo sensing signals in MBS side of the $n$-th subcarrier during the $m$-th OFDM symbol time are~\cite{chen2024downlink}
\begin{equation}  \label{eq2}
\mathbf{y}_{n,m}^\text{MBS}=\sqrt{P_{m,n,\text{T}}^2}\sum_{l=1}^L\left[
\begin{array}{l}
 b_2^le^{j2\pi f_{\text{d},l}^2 mT^2}e^{-j2\pi nQ\Delta f^1 \tau_l}  \\
\times\textbf{a}_\text{Rx}\left(\theta_l\right)\textbf{a}_\text{Tx}^\text{T}\left(\phi_l\right)\text{x}_{n,m}^2
\end{array}\right]+ \text{z}_{n,m}^2,
\end{equation}
where $\mathbf{y}_{n,m}^\text{MBS} \in \mathbb{C}^{N_\text{R}\times 1}$ and $b_2^l$ is the attenuation of the $i$-th target; 
$f_{\text{d},l}^2=\frac{-v_lf_\text{c}^2}{c_0}[\text{cos}(\theta_\text{real}^l-\theta_l)+\text{cos}(\theta_\text{real}^l-\phi_l)]$ is the Doppler frequency shift, while $v_l$ and $\theta_\text{real}^l$ are the magnitude and angle of velocity of the $l$-th target, respectively~\cite{wei2024integrated}; $c_0$ is the speed of light;
$\tau_l=\frac{R_l}{c_0}$ is the delay with $R_l$ is the sum of distances from the $l$-th target to MBS and MiBS;
$\theta_l$ and $\phi_l$ are the angle of arrive (AoA) of the signal from the $l$-th target to MBS and the angle of departure (AoD) of the signal from MiBS to the $l$-th target, respectively;
$\text{x}_{n,m}^2 \in \mathbb{C}^{N_\text{T}\times1}$ and $ \text{z}_{n,m}^2 \sim \mathcal{CN}(0,\sigma_\text{MBS}^2)$ are a modulation symbol vector transmitted by MiBS and an additive white Gaussian noise (AWGN) vector, respectively;
$\textbf{a}_\text{Rx}(\cdot)$ and $\textbf{a}_\text{Tx}(\cdot)$ are the receive and transmit steering vectors, respectively, denoted by~\cite{chen2024downlink}
\begin{equation}\label{eq3}
    \textbf{a}_\text{Rx}(\cdot)=\left[e^{j2\pi k(\frac{d_r}{\lambda})\sin(\cdot)}\right]^\text{T}|_{k=1,2,\cdots,N_\text{R}},
\end{equation}
\begin{equation}\label{eq4}
    \textbf{a}_\text{Tx}(\cdot)=\left[e^{j2\pi p(\frac{d_r}{\lambda})\sin(\cdot)}\right]^\text{T}|_{p=1,2,\cdots,N_\text{T}},
\end{equation}
where $k$ and $p$ are the indices of received and transmit antenna, respectively; $(\cdot)^\text{T}$ is the transpose operator. The space of antennas in both MBS and MiBS is set to half the wavelength for beamforming, namely, $\frac{d_r}{\lambda}=\frac{1}{2}$.

Based on Figs.~\ref{fig1} and (\ref{eq2}), the distance from MBS to targets to MiBS and the distance from MiBS to targets to MBS are identical. In the context of an exceedingly short sensing frame, it is reasonable to maintain that both the AoA and AoD remain unchanged. Therefore, the received echo sensing signals in MiBS side of the $n$-th subcarrier during the $m$-th OFDM symbol time can be expressed as~\cite{chen2024downlink}
\begin{equation}\label{eq5}  \mathbf{y}_{n,m}^\text{MiBS}=\sqrt{P_\text{T}^1}\sum_{l=1}^L\left[
\begin{array}{l}
 b_1^le^{j2\pi f_{\text{d},l}^1 mT^1}e^{-j2\pi n\Delta f^1 \tau_l}  \\
\times\textbf{a}_\text{Rx}\left(\phi_l\right)\textbf{a}_\text{Tx}^\text{T}\left(\theta_l\right)\text{x}_{n,m}^1
\end{array}\right]+ \text{z}_{n,m}^1,
\end{equation}
where $\mathbf{y}_{n,m}^\text{MiBS} \in \mathbb{C}^{N_\text{R}\times 1}$ and $f_{\text{d},l}^1=\frac{-v_lf_\text{c}^1}{c_0}[\text{cos}(\theta_\text{real}^l-\phi_l)+\text{cos}(\theta_\text{real}^l-\theta_l)]$ is the Doppler frequency shift; 
$\text{x}_{n,m}^1 \in \mathbb{C}^{N_\text{T}\times1}$ and $ \text{z}_{n,m}^1 \sim \mathcal{CN}(0,\sigma_\text{MiBS}^2)$ are a modulation symbol vector transmitted by MBS and an AWGN vector, respectively.

\begin{figure*}
    \centering
    \includegraphics[width=0.8\textwidth]{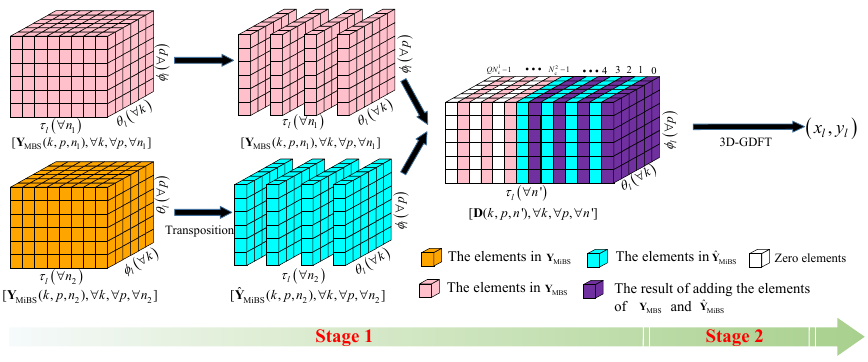}
    \caption{The processing flowchart of the proposed passive sensing echo fusion localization method with $Q$=2}
    \label{fig2}
\end{figure*}

\section{Symbol-Level Fusion Enabled MBS and MiBS Cooperative Passive Sensing Scheme}\label{se3}
The proposed scheme involves two stages: passive sensing data fusion stage and sensing processing stage. 
In the data fusion stage of passive sensing, the passive echo sensing signals received by MBS and MiBS are initially canceled of communication symbols and then fused at the symbol-level.
In the sensing processing stage, we propose a 3D-GDFT method to achieve high-accuracy localization of multiple targets.

\subsection{Passive Sensing Data Fusion Stage}

Firstly, the MBS and MiBS mutually possess each other communication information, enabling the canceling of the communication data in $\mathbf{y}_{n,m}^\text{MBS}$ and $\mathbf{y}_{n,m}^\text{MiBS}$. Without consider the noise, the received echo sensing signals in MBS and MiBS after canceling the communication data can be expressed as (\ref{eq6}) and (\ref{eq7}), respectively.
\begin{equation} \label{eq6}
\begin{aligned} \Hat{\mathbf{y}}_{n,m}^\text{MBS}&=\mathbf{y}_{n,m}^\text{MBS}(\mathbf{x}_{n,m}^2)^\text{H}\left(\mathbf{x}_{n,m}^2(\mathbf{x}_{n,m}^2)^\text{H}\right)^{-1}\\&=  \sqrt{P_\text{T}^2}\sum_{l=1}^L\left[
\begin{array}{l}
 b_2^le^{j2\pi f_{\text{d},l}^2 mT^2}e^{-j2\pi nQ\Delta f^1 \tau_l}  \\
\times\textbf{a}_\text{Rx}\left(\theta_l\right)\textbf{a}_\text{Tx}^\text{T}\left(\phi_l\right)\mathbf{I}_{N_\text{T}}
\end{array}\right],  
\end{aligned}
\end{equation}
\begin{equation} \label{eq7}
\begin{aligned} \Hat{\mathbf{y}}_{n,m}^\text{MiBS}&=\mathbf{y}_{n,m}^\text{MiBS}(\mathbf{x}_{n,m}^2)^\text{H}\left(\mathbf{x}_{n,m}^2(\mathbf{x}_{n,m}^2)^\text{H}\right)^{-1}\\&=  \sqrt{P_\text{T}^1}\sum_{l=1}^L\left[
\begin{array}{l}
 b_1^le^{j2\pi f_{\text{d},l}^1 mT^1}e^{-j2\pi n\Delta f^1 \tau_l}  \\
\times\textbf{a}_\text{Rx}\left(\phi_l\right)\textbf{a}_\text{Tx}^\text{T}\left(\theta_l\right)\mathbf{I}_{N_\text{T}}
\end{array}\right],  
\end{aligned}
\end{equation}
where $\Hat{\mathbf{y}}_{n,m}^\text{MBS} \in \mathbb{C}^{N_\text{R}\times N_\text{T}}$ and $\Hat{\mathbf{y}}_{n,m}^\text{MiBS} \in \mathbb{C}^{N_\text{R}\times N_\text{T}}$;
$(\cdot)^\text{H}$ and $(\cdot)^{-1}$ are conjugate transpose operator and inverse operator, respectively; $\mathbf{I}_{N_\text{T}} \in \mathbb{C}^{N_\text{T}\times N_\text{T}} $ is an identity matrix.

Observing (\ref{eq6}) and (\ref{eq7}), the challenge of symbol-level fusion is the different parameters of the echo sensing signals in MBS and MiBS. To this end, a symbol-level fusion method is proposed to fuse the passive echo sensing signals received by both MBS and MiBS.

According to \cite{sturm2011waveform}, (\ref{eq6}), and (\ref{eq7}), the delay leads to a linear phase shift only along the frequency axis, indicating that one OFDM symbol is sufficient to estimate delay.
Therefore, the received echo sensing signals in MBS and MiBS
after canceling the communication data by the $k$-th receive antenna of the $p$-th transmit antenna on all subcarriers during the $0$-th OFDM symbol time are (\ref{eq8}) and (\ref{eq9}), respectively.
\begin{equation}\label{eq8}
    \mathbf{Y}_{\text{MBS}}(k,p,n_1)=\sqrt{P_\text{T}^2}\sum_{l=1}^Lb_2^l\left[
\begin{array}{l}
e^{-j2\pi n_1Q\Delta f^1 \tau_l}e^{j\pi k\sin{\theta_l}}  \\
\times e^{j\pi p\sin{\phi_l}}
\end{array}\right],
\end{equation}
\begin{equation}\label{eq9}
    \mathbf{Y}_{\text{MiBS}}(k,p,n_2)=\sqrt{P_\text{T}^1}\sum_{l=1}^Lb_1^l\left[
\begin{array}{l}
e^{-j2\pi n_2\Delta f^1 \tau_l}e^{j\pi k\sin{\phi_l}}  \\
\times e^{j\pi p\sin{\theta_l}}
\end{array}\right],
\end{equation}
where $\mathbf{Y}_{\text{MBS}} \in \mathbb{C}^{N_\text{R}\times N_\text{T} \times N_\text{c}^2}$ and $\mathbf{Y}_{\text{MiBS}} \in \mathbb{C}^{N_\text{R}\times N_\text{T} \times N_\text{c}^1}$;
$n_1\in \{0,1,\cdots,N_\text{c}^2\}$ and $n_2\in \{0,1,\cdots,N_\text{c}^1\}$.

\begin{table}[!ht]
\centering
\label{tab2}
\resizebox{0.42\textwidth}{!}{
\begin{tabular}{rllll}
\hline
\multicolumn{5}{l}{\textbf{Algorithm 1:} Target localization method}   \\ \hline
\multirow{-5}{*}{\textbf{Input:} }               & \multicolumn{4}{l}{\begin{tabular}[c]{@{}l@{}} The received echo sensing signals in MBS and MiBS \\ after stripping the communication data $\mathbf{Y}_\text{MBS}$ and $\mathbf{Y}_\text{MiBS}$;\\ The grid position vector $\mathbf{C}$;\\
The ratio of subcarrier spacing $Q$; \\ The locations of MBS and MiBS $(x_\text{M},y_\text{M})$ and $(x_\text{Mi},y_\text{Mi})$. \end{tabular}} \\
\textbf{Output:}               & \multicolumn{4}{l}{The locations of $L$ targets $\{(x_1,y_1),(x_2,y_2),\cdots,(x_L,y_L)\}$.} \\ 
\textbf{Stage 1:}  & \multicolumn{4}{l}{\textbf{Fusing the echo sensing data in MBS and MiBS.}} \\
1:      & \multicolumn{4}{l}{Initialize a 3D zero matrix $\mathbf{D}$ and an index value $n_1\in \mathcal{N}_1$;} \\
2:      & \multicolumn{4}{l}{Transpose $\mathbf{Y}_\text{MiBS}$ to obtain $\widehat{\mathbf{Y}}_{\text{MiBS}}$;}   \\
3:      & \multicolumn{4}{l}{\textbf{For} $\mathbf{D}$ ($n_1$ in $\mathcal{N}_1$) $\textbf{do}$}   \\
4:        & \multicolumn{4}{l}{$\hspace{1em}$ The element values of $\widehat{\mathbf{Y}}_{\text{MiBS}}(:,:,n_1)$ are assigned to $\mathbf{D}(:,:,n_1)$;}  \\
5:        & \multicolumn{4}{l}{\textbf{End For}}  \\
6:        & \multicolumn{4}{l}{Initialize an index value $n_2 \in \mathcal{N}_2$;}  \\
7:       & \multicolumn{4}{l}{ $\textbf{For}$ $\mathbf{D}$ $(n_2$ in $\mathcal{N}_2)$ $\textbf{do}$}   \\
8:       & \multicolumn{4}{l}{$\hspace{1em}$ The element values of $\mathbf{Y}_\text{MBS}(:,:,n_2)$ are added to $\mathbf{D}(:,:,Qn_2)$;}   \\
9:       & \multicolumn{4}{l}{$\hspace{1em}$ $\textbf{If}$ $Qn_2 \le N_\text{c}^2-1$ $\textbf{do}$}  \\
10:       & \multicolumn{4}{l}{$\hspace{2em}$ $\mathbf{D}(:,:,Qn_2)$ = $\mathbf{D}(:,:,Qn_2)/(2N^2)$;}  \\
11:       & \multicolumn{4}{l}{$\hspace{1em}$ \textbf{End If}}  \\
12:      & \multicolumn{4}{l}{\textbf{End For}}  \\
\textbf{Stage 2:}  & \multicolumn{4}{l}{\textbf{Targets localization with a 3D-GDFT method}.} \\
1:      & \multicolumn{4}{l}{ According to $\mathbf{C}$, $(x_\text{M},y_\text{M})$, and $(x_\text{Mi},y_\text{Mi})$, the AoA, AoD, and  } \\
& \multicolumn{4}{l}{delay searching vectors are obtained, denoted by $\mathbf{E}$, $\mathbf{F}$, and $\mathbf{G}$;}\\
2:      & \multicolumn{4}{l}{Based on $\mathbf{E}$, $\mathbf{F}$, and $\mathbf{G}$ to construct $G$ matching matrices $\mathbf{H}$;}\\
3:      & \multicolumn{4}{l}{The Hadamard products of $G$ matching matrices $\mathbf{H}$ and $\mathbf{D}$ are }\\  & \multicolumn{4}{l}{performed to obtain a searching spectrum vector $\mathbf{P}$;}  \\
4:     & \multicolumn{4}{l}{Search the peaks of $\mathbf{P}$ to obtain peak indices $\{\xi_1,\xi_2,\cdots,\xi_L\}$;}\\ 
5:     & \multicolumn{4}{l}{The location of targets are obtained by $\mathbf{C}(\{\xi_1,\xi_2,\cdots,\xi_L\})$.}  \\
\hline
\end{tabular}}
\end{table}

Without loss of generality, we assume that $N_\text{T}=N_\text{R}=N$ and the phase of $b_i^l$ is small. When the subcarrier index is fixed, the $\theta_l$ and $\phi_l$ are along the receive and transmit antenna dimension of $\mathbf{Y}_{\text{MBS}}$, respectively, which is reversed in $\mathbf{Y}_{\text{MiBS}}$. Meanwhile, the size of the phase shift introduced by the $\tau_l$ along the subcarrier dimension of $\mathbf{Y}_{\text{MBS}}$ is $Q$ times as large as that in $\mathbf{Y}_{\text{MiBS}}$. Therefore, the steps to fuse $\mathbf{Y}_{\text{MBS}}$ and $\mathbf{Y}_{\text{MBS}}$ are as follows.
\begin{itemize}
    \item \textbf{Step 1:} Constructe a three-dimensional (3D) zero matrix $\mathbf{D} \in \mathbb{C}^{N \times N\times (QN_\text{c}^1-1)}$ and an index value $n' \in \{0,1,\cdots,QN_\text{c}^1-1\}$.
    \item \textbf{Step 2:} Transpose the transmit and receive antenna dimensions of $\mathbf{Y}_{\text{MiBS}}$ to obtain the transposed matrix $\widehat{\mathbf{Y}}_{\text{MiBS}}$, where $\widehat{\mathbf{Y}}_{\text{MiBS}}(k,p,n_2)=\mathbf{Y}_{\text{MiBS}}(p,k,n_2)$.
    \item \textbf{Step 3:} Traverse index values 
$k$, $p$, and $n'$ in $\{1,2,\cdots,N\}$, $\{1,2,\cdots,N\}$, and $\{0,1,\cdots,QN_\text{c}^1-1\}$, respectively. When $n'=n_2 \ne Qn_1$, $\mathbf{D}(k,p,n')=\widehat{\mathbf{Y}}_{\text{MiBS}}(k,p,n_2)$. When $n'=n_2 = Qn_1$, $\mathbf{D}(k,p,n')=\frac{1}{2N\times N}\left[\widehat{\mathbf{Y}}_{\text{MiBS}}(k,p,n_2)+\mathbf{Y}_{\text{MBS}}(k,p,n_1)\right]$. When $n'= Qn_1>(N_\text{c}^2-1)$, $\mathbf{D}(k,p,n')=\mathbf{Y}_{\text{MBS}}(k,p,n_1)$.
\end{itemize}

\subsection{Sensing Processing Stage}
Observing the matrix $\mathbf{D}$, the linear phase offsets introduced by AoA, AoD, and delay are orthogonal. For single-dimensional estimation, 3D processing gains cannot be obtained. Regarding the multi-dimensional joint estimation, the 3D multiple signal classification (3D-MUSIC) method~\cite{zhao2019three} requires the number of targets, and the 3D-DFT method~\cite{lu2023isac} is constrained by bandwidth and space-domain resources.

Therefore, a 3D-GDFT method is proposed to obtain 3D gains and high-accuracy sensing without requiring the number of targets and enormous bandwidth resources. The steps of the proposed 3D-GDFT method are as follows.
\begin{itemize}
    \item \textbf{Step 1:} Use the coverage area of MiBS as the searching scope and grid it to obtain grid position vector, which is denoted by
    \begin{equation}\label{eq10}
        \mathbf{C}=[(x_\text{grid}^1,y_\text{grid}^1),\cdots,(x_\text{grid}^g,y_\text{grid}^g),\cdots,(x_\text{grid}^G,y_\text{grid}^G)],
    \end{equation}
    where $g\in\{1,2,\cdots,G\}$ and $G$ are the index and number of grids, respectively.  
    \item \textbf{Step 2:} According to $\mathbf{C}$, the location $(x_\text{M},y_\text{M})$ of MBS, and the location $(x_\text{Mi},y_\text{Mi})$ of MiBS, the corresponding AoA and AoD searching vectors can be expressed as (\ref{eq11}) and (\ref{eq12}), respectively.
    \begin{equation}\label{eq11}
        \mathbf{E}=[\theta_\text{grid}^1,\cdots,\theta_\text{grid}^g,\cdots,\theta_\text{grid}^G],
    \end{equation}
    \begin{equation}\label{eq12}
        \mathbf{F}=[\phi_\text{grid}^1,\cdots,\phi_\text{grid}^g,\cdots,\phi_\text{grid}^G].
    \end{equation}
    \item \textbf{Step 3:} Based on $\mathbf{C}$, we calculate a delay searching vector, which is denoted by
    \begin{equation} \label{eq13}
        \mathbf{G}=[\tau_\text{grid}^1,\cdots,\tau_\text{grid}^g,\cdots,\tau_\text{grid}^G],
    \end{equation}
    where 
     {\tiny\begin{equation*}
        \tau_\text{grid}^g = \sqrt{(x_\text{M}-x_\text{grid}^g)^2+(y_\text{M}-y_\text{grid}^g)^2}+\sqrt{(x_\text{Mi}-x_\text{grid}^g)^2+(y_\text{Mi}-y_\text{grid}^g)^2}.
    \end{equation*}}
    
    \item \textbf{Step 4:} Construct $G$ matching matrices $\mathbf{H}\in \mathbb{C}^{N\times N \times N_{\text{c}}^1}$ based on $\mathbf{G}$, $\mathbf{E}$, and $\mathbf{F}$, the $(k,p,n')$-th element of the $g$-th matching matrix $\mathbf{H}^g$ is 
    \begin{equation} \label{eq14}
        \mathbf{H}_{k,p,n'}^g=e^{-j\pi k\sin \theta_\text{grid}^g}
        e^{-j\pi p\sin \phi_\text{grid}^g}e^{j2\pi n'\Delta f^1 \tau_\text{grid}^g}.  
    \end{equation}
    \item \textbf{Step 5:} The Hadamard products of $G$ matrix $\mathbf{H}$ and $\mathbf{D}$ are performed, yielding $G$ matrices $\mathbf{\hat{H}}$, where the $g$-th matrix $\mathbf{\hat{H}}^g$ is represented as
    \begin{equation} \label{eq15}
        \mathbf{\hat{H}}^g=\mathbf{H}^g \circ \mathbf{D},
    \end{equation}
    where $\circ$ is the Hadamard product.
    \item \textbf{Step 6:} Obtain a searching spectrum vector $\mathbf{P} \in \mathbb{C}^{1 \times G}$, the $g$-th element of $\mathbf{P}$ is
    \begin{equation}\label{eq16}
        \mathbf{P}(g)=|\sum_{k=1}^N\sum_{p=1}^N\sum_{n'=0}^{QN_{\text{c}}^1-1}\mathbf{\hat{H}}^g(k,p,n')|,
    \end{equation}
    where $|\cdot|$ is absolute value operator. 
    \item \textbf{Step 7:} Search $\mathbf{P}$ and obtain the $L$ peak indices $\{\xi_1,\cdots,\xi_l,\cdots,\xi_L\}$, the estimated location of the $l$-th target is $\mathbf{C}(\xi_l)$.
\end{itemize}


The proposed target localization method is shown in Algorithm 1, which is explained intuitively in Fig.~\ref{fig2}.

\begin{table*}[!htbp]
	\caption{ Simulation parameters \cite{access2009base,wei2023carrier,muhammed2020energy}}
	\label{tab_simulation}
	\renewcommand{\arraystretch}{1.3} 
	\begin{center}\resizebox{0.85\linewidth}{!}{
		\begin{tabular}{|m{0.07\textwidth}<{\centering}| m{0.32\textwidth}<{\centering}| m{0.13\textwidth}<{\centering}| m{0.1\textwidth}<{\centering}|m{0.32\textwidth}<{\centering}| m{0.13\textwidth}<{\centering}|}
			\hline
			\textbf{Symbol} & \textbf{Parameter} & \textbf{Value} & \textbf{Symbol} & \textbf{Parameter} & \textbf{Value} \\
			\hline
			$N_\text{c}^1$, $N_\text{c}^2$	& Number of subcarriers in MBS and MiBS & $512$ & $M_{\text{sym}}^1$, $M_{\text{sym}}^2$	& Number of OFDM symbols in MBS and MiBS  & $128$  \\
                \hline
			$N_\text{R}$, $N_\text{T}$ & Number of receive and transmit antennas in MBS and MiBS & $64, 64$ & $P_\text{T}^1$, $P_\text{T}^2$ & Transmit power of MBS and MiBS & (46, 27) dBm   \\
			\hline
			$f_\text{c}^1$, $f_\text{c}^2$	& Carrier frequency & (2.6,\ 26) GHz & $\Delta f^1$, $\Delta f^2$	& Subcarrier spacing  & (30,\ 120) kHz \\
			\hline
		 $L$& Number of targets & 3	& $(x_1,y_1)$	& The coordinate of the $1$-st target & (200, 30) m  \\			
			\hline
		  $(x_2,y_2)$ & The coordinate of the $2$-nd target & (250, 60) m & $(x_3,y_3)$	& The coordinate of the $3$-rd target & (300, 80) m  \\
                \hline
         $(x_\text{M},y_\text{M})$ & The coordinate of MBS & (0, 0) m & $(x_{\text{Mi}},y_{\text{Mi}})$	& The coordinate of MiBS & (300, 0) m   \\
			\hline
		\end{tabular}}
	\end{center}
\end{table*}

\section{Simulation Results}\label{se4}
In this section, the proposed target localization method is evaluated. Specifically, the two-dimensional (2D) location map and sum of root mean square error (SMSE) of multiple targets localization are simulated. The simulation parameters are listed in Table~\ref{tab_simulation}. Unless otherwise specified, the independent variable in the simulation is set to the noise power to fairly compare the performance of MBS and MiBS under cooperative and non-cooperative scenarios (``non-cooperative'' refers to the MBS passive sensing or MiBS passive sensing).

\subsection{2D Location Map}
\begin{figure}[!ht]
    \centering
    \includegraphics[width=0.4\textwidth]{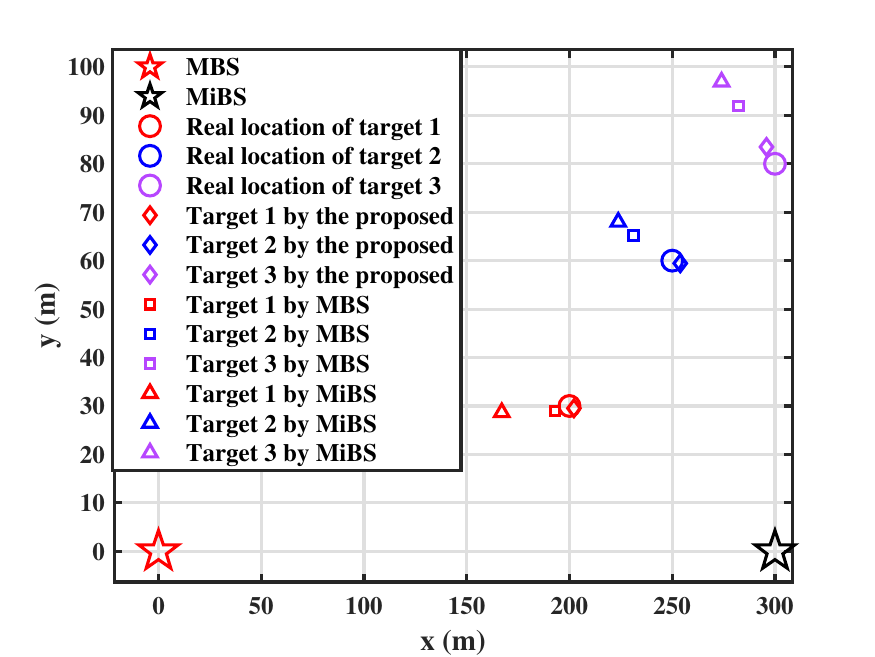}
    \caption{The 2D location map with the power of noise is -200 dBm/Hz~\cite{goldsmith2005wireless}}
    \label{fig3}
\end{figure}

With the proposed 3D-GDFT method, Fig.~\ref{fig3} shows the 2D location map of target localization with MBS and MiBS cooperative and non-cooperative sensing scenarios. According to Fig.~\ref{fig3},
the diamond marks are closer to the real location of targets, reveling superior target localization capabilities of the proposed MBS and MiBS cooperative passive sensing scheme.

\subsection{SMSE of Localization Estimation}
The SMSEs of multiple targets localization are simulated when the scope of noise power is -175 dBm/Hz to -135 dbm/Hz~\cite{goldsmith2005wireless}.
Firstly, the SMSEs of the MBS and MiBS cooperative and non-cooperative passive sensing are simulated to verify the superiority of cooperative passive sensing. Then, the SMSEs of different methods are simulated to evaluate the high-accuracy sensing of the proposed 3D-GDFT method. Finally, the data-level fusion and symbol-level fusion sensing are compared to reveal the high-accuracy performance of symbol-level fusion.

\subsubsection{Cooperative v.s. non-cooperative}
\begin{figure}[htbp]
	\centering
	\subfigure[3D-DFT method] {\label{fig4.a}\includegraphics[width=.24\textwidth]{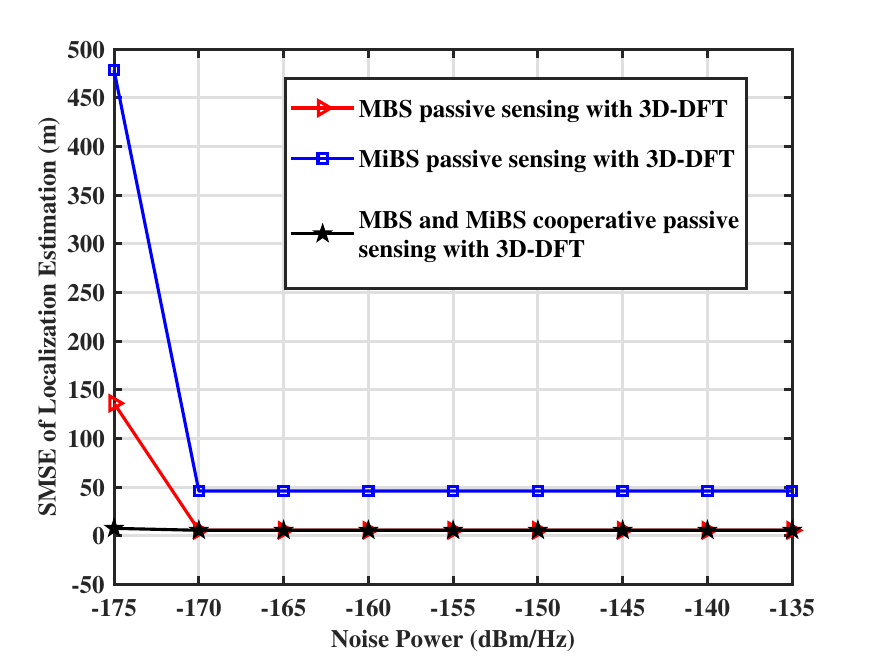}}
	\subfigure[3D-MUSIC method] {\label{fig4.b}\includegraphics[width=.24\textwidth]{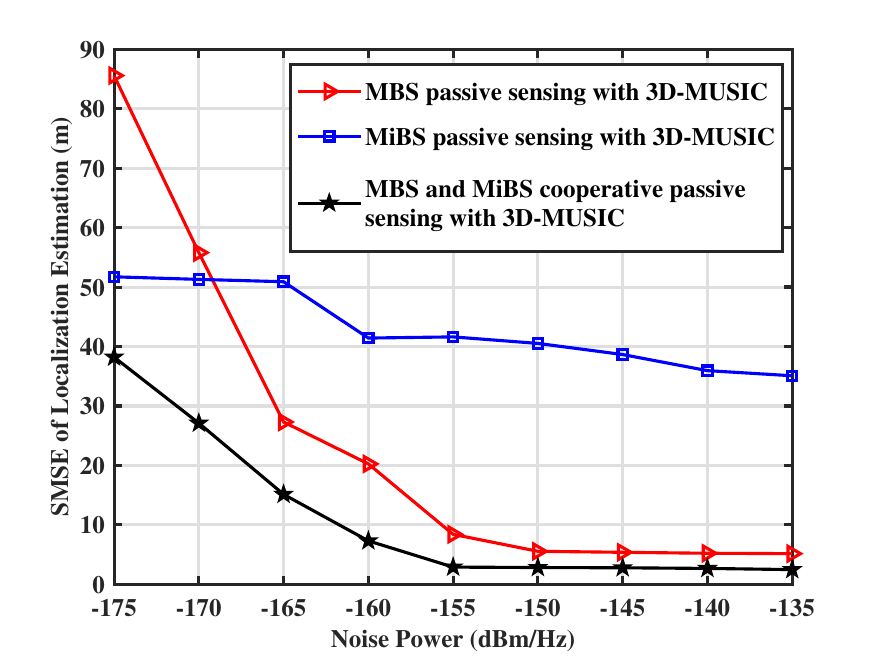}}
        \subfigure[Proposed 3D-GDFT method] {\label{fig4.c}\includegraphics[width=.3\textwidth]{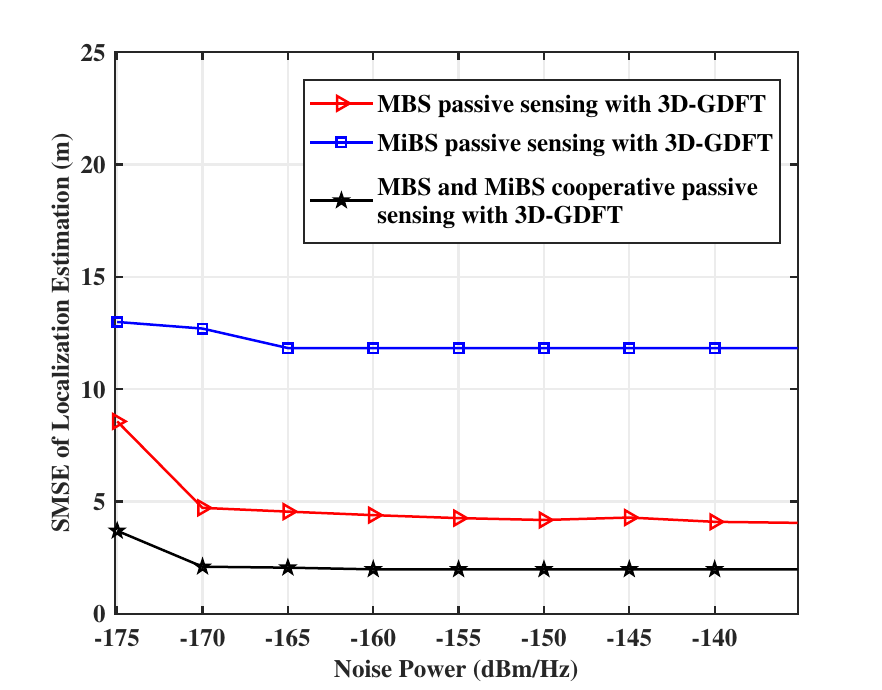}}
	\caption{Cooperative \textit{v.s.} non-cooperative passive sensing with different methods}
	\label{fig.4}
\end{figure}

Fig.~\ref{fig.4} shows the SMSE of cooperative and non-cooperative passive sensing with different methods. 

\textbf{3D-DFT:} The processing for target localization using the 3D-DFT method in simulation is shown in Appendix A. 

\textbf{3D-MUSIC:} The processing is similar to 3D-DFT, with the distinction that it necessitates prior knowledge of the number of targets to extract the noise subspace.

As shown in (a), (b), and (c) of Fig.~\ref{fig.4}, the MiBS passive sensing shows high SMSE due to its narrow bandwidth, while MBS and MiBS cooperative passive sensing exhibits the lowest SMSE, highlighting the superior sensing performance of the proposed MBS and MiBS cooperative passive sensing scheme.

\subsubsection{Comparison of 3D-MUSIC, 3D-DFT, and 3D-GDFT methods}
\begin{figure}[htbp]
	\centering
	\subfigure[MiBS passive sensing] {\label{fig5.a}\includegraphics[width=.24\textwidth]{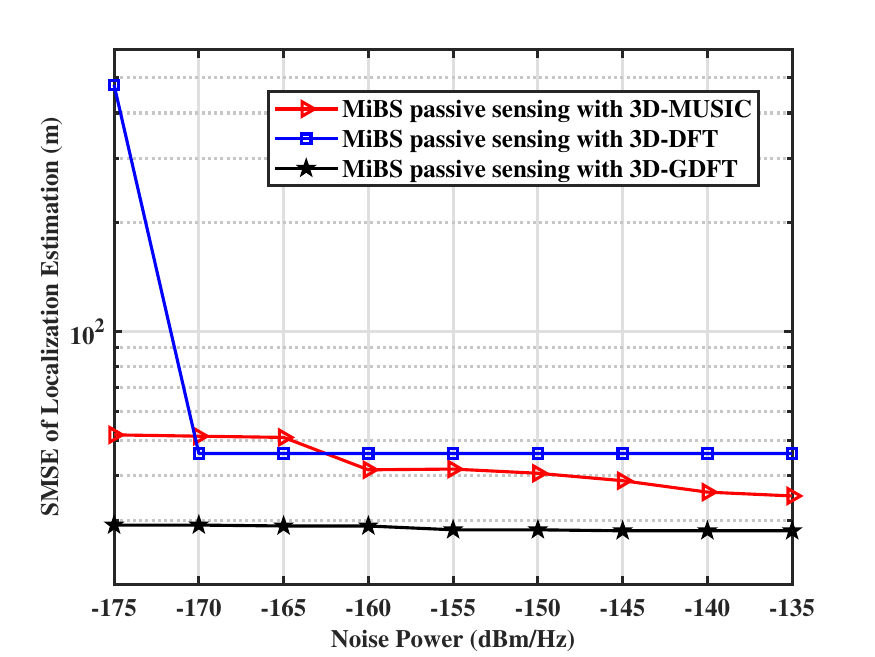}}
	\subfigure[MBS passive sensing] {\label{fig5.b}\includegraphics[width=.24\textwidth]{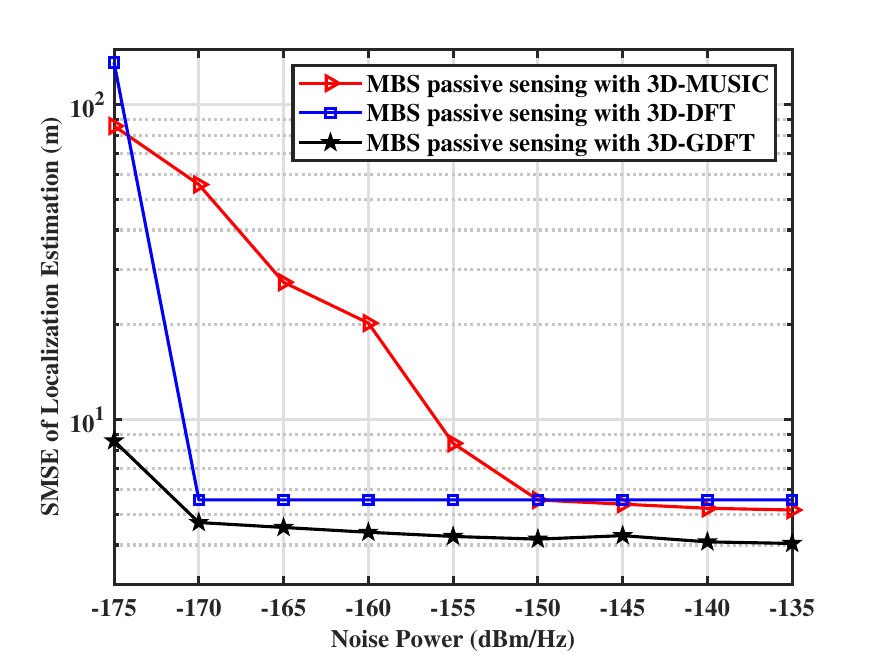}}
        \subfigure[MBS and MiBS cooperative passive sensing] {\label{fig5.c}\includegraphics[width=.3\textwidth]{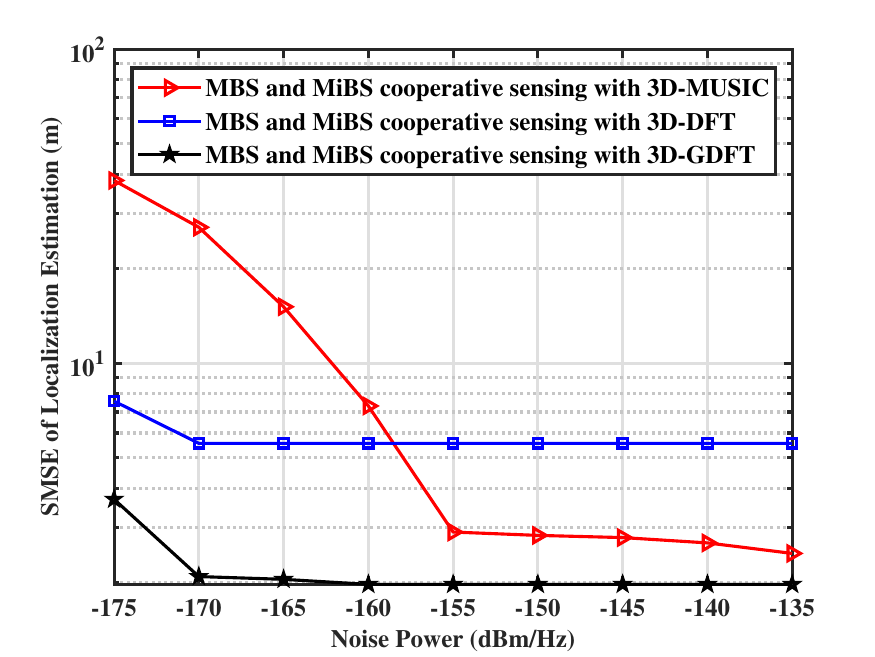}}
	\caption{Comparison of different methods with different scenarios}
	\label{fig.5}
\end{figure}

As shown in (a), (b), and (c) of Fig.~\ref{fig.5}, the SMSEs of 3D-GDFT with different scenarios are the lowest, revealing the high-accuracy localization of the proposed 3D-GDFT method. Meanwhile, when the noise power is high, the red line is higher than the blue line, because the 3D-MUSIC method is hardly capable to obtain the entire 3D processing gains. 

Another phenomenon is that the red line and the black line tend to flat as the noise power decreases, because the 3D-DFT method has a theoretical lower bound due to resolution limitations~\cite{liu2023isac}, while the proposed 3D-GDFT method is primarily constrained by grid size. It should be noted that the complexity of the 3D-MUSIC, 3D-DFT, and proposed 3D-GDFT methods are
$\mathcal{O}\left\{N_\text{R}^3+N_\text{R}^2+\Psi^2(N_\text{R}+1)(N_\text{R}-1)+(N_\text{c}^1)^3+(N_\text{c}^1)^2+\Phi^2\right.$ $\left.(N_\text{c}^1+1)(N_\text{c}^1-1)\right\}$, $\mathcal{O}\left\{N_\text{R}^2+(N_\text{c}^1)^2\right\}$, and $G^2(N_\text{R}^2N_\text{c}^1Q)$, respectively, where $\Phi$ and $\Psi$ are the numbers of search grid. The complexity of the proposed method is higher than the other two, which indicates that the proposed method sacrifices some efficiency.

\subsubsection{Data-level fusion v.s. symbol-level fusion}
\begin{figure}
    \centering
    \includegraphics[width=0.3\textwidth]{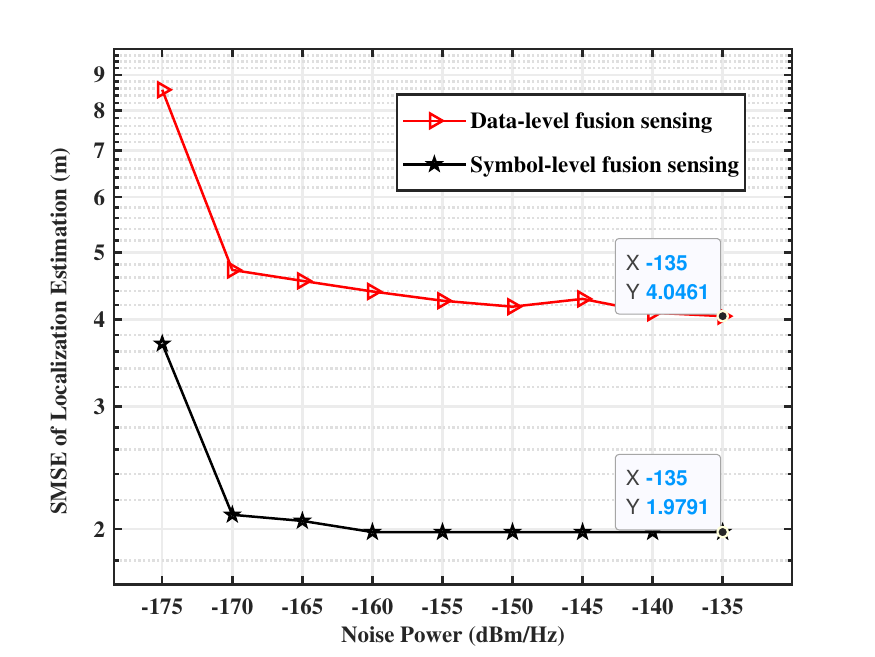}
    \caption{Data-level fusion \textit{v.s.} symbol-level fusion}
    \label{fig6}
\end{figure}

In simulation, data-level fusion refers to MBS and MiBS independently processing received echo sensing signals to obtain location information for multiple targets and subsequently averaging the results.
As depicted in Fig.~\ref{fig6}, the symbol-level fusion achieves localization accuracy one order of magnitude higher than data-level fusion, enabling meter-level localization accuracy with limited resources.

\section{Conclusion}\label{se5}
With the coexistence and overlapped coverage of MBS and MiBS, MBS and MiBS cooperative sensing has become a widespread research topic. In this paper, we consider MBS and MiBS cooperative passive sensing for multiple targets localization in IoV scenario. A symbol-level fusion method and a 3D-GDFT method are proposed to achieve high-accuracy localization with limited resources. Simulation results demonstrate the superior localization performance of the proposed MBS and MiBS cooperative passive sensing scheme.

\begin{appendices}
\section{3D-DFT for localization} \label{apdA}
For $\mathbf{D}$, the linear phase offsets introduced by AoA, AoD, and delay are orthogonal. Therefore, DFT, DFT, and IDFT are performed sequentially on the AoA, AoD, and delay dimensions of $\mathbf{D}$, yielding processing gains of approximately $N^2(N_\text{c}^1+N_\text{c}^2)$. 
Then, the 3D indices of the $l$-th target are obtained, denoted as $\alpha_l$, $\beta_l$, and $\gamma_l$, respectively.

According to \cite{sturm2011waveform}, the 3D indices are transformed to estimated values $\hat{\theta}_l$, $\hat{\phi}_l$, and $\hat{\tau}_l$, respectively. 
\begin{equation}
 \hat{\theta}_l=\arcsin{\frac{(\alpha_l-1)\lambda}{d_\text{r}N}}, 
\end{equation}
\begin{equation}
 \hat{\phi}_l=\arcsin{\frac{(\beta_l-1)\lambda}{d_\text{r}N}}, 
\end{equation}
\begin{equation}
 \hat{\tau}_l=\frac{(\gamma_l-1)}{\Delta f^1QNN_\text{c}^1}.
\end{equation}

Then, the traditional AoA localization method is used to estimate location, with the details are as follows.

Based on the location $(x_\text{M},y_\text{M})$ of MBS, the location $(x_\text{Mi},y_\text{Mi})$ of MiBS, $\hat{\theta_l}$, and $\hat{\tau_l}$, the convectional AoA location algorithm~\cite{xu2008aoa} is performed to obtain the locations of $L$ targets, where the location of $l$-th target is expressed as
     \begin{equation}\label{ap1}
      \begin{cases} x_l=x_\text{M}+d_{\text{ro}}\cos(\hat{\theta_l}),
     \\y_l=y_\text{M}+d_{\text{ro}}\sin(\hat{\theta_l}),
      \end{cases}
      \end{equation} and
     \begin{equation} \label{eqap2}
         d_{\text{ro}}=\frac{(\hat{\tau_l}c_0)^2-(x_\text{M}-x_\text{Mi})^2+(y_\text{M}-y_\text{Mi})^2}{2\left[\hat{\tau_l}c_0-\sqrt{(x_\text{M}-x_\text{Mi})^2+(y_\text{M}-y_\text{Mi})^2}\cos(\hat{\theta_l})\right]}.
     \end{equation}

\end{appendices}

\section*{Acknowledgment}
This work was supported in part by the National Natural Science Foundation of
China (NSFC) under Grant 62271081, in
part by the Fundamental Research Funds for the Central
Universities under Grant 2024ZCJH01, 
and in part by the
National Key Research and Development Program
of China under Grant 2020YFA0711302.

\bibliographystyle{IEEEtran}
\bibliography{reference}

\begin{thebibliography}{10}
\providecommand{\url}[1]{#1}
\csname url@samestyle\endcsname
\providecommand{\newblock}{\relax}
\providecommand{\bibinfo}[2]{#2}
\providecommand{\BIBentrySTDinterwordspacing}{\spaceskip=0pt\relax}
\providecommand{\BIBentryALTinterwordstretchfactor}{4}
\providecommand{\BIBentryALTinterwordspacing}{\spaceskip=\fontdimen2\font plus
\BIBentryALTinterwordstretchfactor\fontdimen3\font minus \fontdimen4\font\relax}
\providecommand{\BIBforeignlanguage}[2]{{%
\expandafter\ifx\csname l@#1\endcsname\relax
\typeout{** WARNING: IEEEtran.bst: No hyphenation pattern has been}%
\typeout{** loaded for the language `#1'. Using the pattern for}%
\typeout{** the default language instead.}%
\else
\language=\csname l@#1\endcsname
\fi
#2}}
\providecommand{\BIBdecl}{\relax}
\BIBdecl

\bibitem{wei2023multiple}
Z.~Wei, F.~Li, H.~Liu, X.~Chen, H.~Wu, K.~Han, and Z.~Feng, ``{Multiple reference signals collaborative sensing for integrated sensing and communication system towards 5G-A and 6G},'' \emph{IEEE Transactions on Vehicular Technology}, 2024.

\bibitem{amhaz2023integrated}
A.~Amhaz, M.~Elhattab, C.~Assi, and S.~Sharafeddine, ``{Integrated Sensing and Communication: NOMA vs Cooperative NOMA},'' in \emph{GLOBECOM 2023-2023 IEEE Global Communications Conference}.\hskip 1em plus 0.5em minus 0.4em\relax IEEE, 2023, pp. 407--412.

\bibitem{niu2024interference}
Y.~Niu, Z.~Wei, L.~Wang, H.~Wu, and Z.~Feng, ``{Interference Management for Integrated Sensing and Communication Systems: A Survey},'' \emph{arXiv preprint arXiv:2403.16189}, 2024.

\bibitem{wei2024deep}
Z.~Wei, H.~Liu, Z.~Feng, H.~Wu, F.~Liu, Q.~Zhang, and Y.~Du, ``Deep cooperation in isac system: Resource, node and infrastructure perspectives,'' \emph{IEEE Internet of Things Magazine}, 2024.

\bibitem{xianrong2020research}
W.~Xianrong, Y.~Jianxin, Z.~Weijie, X.~Deqiang, S.~Kan, S.~Jiale, C.~Feng, R.~Yunhua, G.~Ziping, and K.~Hengyu, ``Research progress and development trend of the multi-illuminator-based passive radar,'' \emph{Journal of Radars}, vol.~9, no.~6, pp. 939--958, 2020.

\bibitem{liu20226g}
J.~Liu, Z.~Jiang, Z.~Sun, and W.~Tian, ``{Integration of 6G Sensing and Communication for Network Fusion},'' \emph{Mobile Communications}, 2022.

\bibitem{wei2023symbol}
Z.~Wei, R.~Xu, Z.~Feng, H.~Wu, N.~Zhang, W.~Jiang, and X.~Yang, ``{Symbol-level integrated sensing and communication enabled multiple base stations cooperative sensing},'' \emph{IEEE Transactions on Vehicular Technology}, 2023.

\bibitem{liu2023isac}
H.~Liu, Z.~Wei, F.~Li, Y.~Lin, H.~Qu, H.~Wu, and Z.~Feng, ``{Integrated Sensing and Communication Signal Processing Based On Compressed Sensing Over Unlicensed Spectrum Bands},'' \emph{IEEE Transactions on Cognitive Communications and Networking}, pp. 1--1, 2024.

\bibitem{chen2024downlink}
X.~Chen, Z.~Feng, J.~A. Zhang, Z.~Wei, X.~Yuan, P.~Zhang, and J.~Peng, ``{Downlink and Uplink Cooperative Joint Communication and Sensing},'' \emph{IEEE Transactions on Vehicular Technology}, 2024.

\bibitem{wei2024integrated}
Z.~Wei, H.~Liu, H.~Li, W.~Jiang, Z.~Feng, H.~Wu, and P.~Zhang, ``{Integrated Sensing and Communication Enabled Cooperative Passive Sensing Using Mobile Communication System},'' \emph{arXiv preprint arXiv:2405.09179}, 2024.

\bibitem{sturm2011waveform}
C.~Sturm and W.~Wiesbeck, ``Waveform design and signal processing aspects for fusion of wireless communications and radar sensing,'' \emph{Proceedings of the IEEE}, vol.~99, no.~7, pp. 1236--1259, 2011.

\bibitem{zhao2019three}
E.~Zhao, F.~Zhang, D.~Zhang, and S.~Pan, ``{Three-dimensional multiple signal classification (3D-MUSIC) for super-resolution FMCW radar detection},'' in \emph{2019 IEEE MTT-S International Wireless Symposium (IWS)}.\hskip 1em plus 0.5em minus 0.4em\relax IEEE, 2019, pp. 1--3.

\bibitem{lu2023isac}
B.~Lu, Z.~Wei, L.~Wang, R.~Zhang, and Z.~Feng, ``{ISAC 4D Imaging System Based on 5G Downlink Millimeter Wave Signal},'' in \emph{2023 IEEE Globecom Workshops (GC Wkshps)}, 2023, pp. 389--394.

\bibitem{access2009base}
E.~U. T.~R. Access, ``{Base Station (BS) radio transmission and reception},'' \emph{3GPP TS}, vol.~36, p.~V9, 2009.

\bibitem{wei2023carrier}
Z.~Wei, H.~Liu, X.~Yang, W.~Jiang, H.~Wu, X.~Li, and Z.~Feng, ``{Carrier Aggregation Enabled Integrated Sensing and Communication Signal Design and Processing},'' \emph{IEEE Transactions on Vehicular Technology}, 2023.

\bibitem{muhammed2020energy}
A.~J. Muhammed, Z.~Ma, Z.~Zhang, P.~Fan, and E.~G. Larsson, ``{Energy-efficient resource allocation for NOMA based small cell networks with wireless backhauls},'' \emph{IEEE Transactions on Communications}, vol.~68, no.~6, pp. 3766--3781, 2020.

\bibitem{goldsmith2005wireless}
A.~Goldsmith, \emph{Wireless communications}.\hskip 1em plus 0.5em minus 0.4em\relax Cambridge university press, 2005.

\bibitem{xu2008aoa}
J.~Xu, M.~Ma, and C.~L. Law, ``{AOA cooperative position localization},'' in \emph{IEEE GLOBECOM 2008-2008 IEEE Global Telecommunications Conference}.\hskip 1em plus 0.5em minus 0.4em\relax IEEE, 2008, pp. 1--5.

\end{thebibliography}

\end{document}